\documentclass[aps,pra,twocolumn,superscriptaddress]{revtex4-2}

\usepackage{amsmath,amssymb}
\usepackage{physics}
\usepackage{braket}
\usepackage{graphicx}
\usepackage{hyperref}
\usepackage{bbold}
\usepackage{xcolor}

\begin{document}

\title{Coherence Response in Noisy Quantum Measurements}

% \title{Coherence-Sensitive Readout Models for Quantum Devices:\\
% Beyond the Classical Assignment Matrix}

% Alternative titles
% 
% Modeling Coherence-induced measurement errors for Quantum Devices: Beyond the Classical Assignement Matrix
%
% A Coherence-Aware Framework for Quantum Readout Error Mitigation 
%
% Coherence-Aware Quantum Measurement Error Mitigation Beyond the Classical Model
% 
% 

\author{Zachariah Malik
}
\address{‡Department of Applied Mathematics, University of Colorado, Boulder, CO 80309-0526, USA}

\author{Quinn Langfitt
}
\address{Department of Computer Science, Northwestern University, Evanston, IL, USA}

\author{Zain H. Saleem
}
\address{Mathematics and Computer Science Division, Argonne National Laboratory, Lemont, IL, USA}

\date{\today}

\begin{abstract}
Readout error models for noisy quantum devices almost universally assume that measurement noise is classical: the measurement statistics are obtained from the ideal computational-basis populations by a column-stochastic assignment matrix $A$. This description is equivalent to assuming that the effective positive-operator-valued measurement (POVM) is diagonal in the measurement basis, and therefore completely insensitive to quantum coherences. We relax this assumption and derive a fully general expression for the observed measurement probabilities under arbitrary completely positive trace-preserving (CPTP) noise preceding a computational-basis measurement. Writing the ideal post-circuit state $\tilde{\rho}$ in terms of its populations $x$ and coherences $y$, we show that the observed probability vector $z$ satisfies $z = A x + C y$, where $A$ is the familiar classical assignment matrix and $C$ is a coherence-response matrix constructed from the off-diagonal matrix elements of the effective POVM in the computational basis. The classical model $z = A x$ arises if and only if all POVM elements are diagonal; in this sense $C$ quantifies accessible information about coherent readout distortions and interference between computational-basis states, all of which are invisible to models that retain only $A$. Our numerical experiments show that incorporating $C$ into readout recovery can improve fidelity over classical inversion and enable selective Pauli twirling with exponentially reduced circuit overhead. This work therefore provides a natural, fully general framework for coherence-sensitive readout modeling on current and future quantum devices.
\end{abstract}

\maketitle

\section{Introduction}
\label{sec:intro}

Recent advances in quantum fabrication technologies have resulted in quantum computing systems with access to thousands of qubits in continuous operation \cite{Chiu2025}, \cite{D-Wave_Whitepaper}. As quantum devices grow in scale, they are finding more applications across various domains. This includes quantum chemistry \cite{Guglielmo2024}, \cite{Motta_2024}, \cite{McArdle2020}, physics \cite{Barthe2025}, \cite{Schuhmacher2025}, finance \cite{Herman2023}, \cite{Ciceri2025}, numerical methods \cite{Shu2024}, \cite{Bochkarev2026}, among others. 

However, noise remains an issue for quantum devices, thus necessitating the use of readout error mitigation techniques \cite{Fellous-Asiani2025}, \cite{Yan2025}, \cite{Quek2024}. In contrast to error correction via quantum codes (see, e.g., \cite{Roffe2019} or \cite[Chapter 10]{NielsenChuang2010} for a review of quantum error codes), the error mitigation techniques we are interested in do not introduce additional overhead in terms of the number of qubits or gates used. 

Let $x, z \in \mathbb{R}^{N}$ be the vectors of ideal and noisy outcome probabilities, respectively, of a quantum device of $n$ qubits, with $N := 2^{n}$, measured in the computational basis. The literature often assumes that there exists a \emph{classical assignment matrix} $A \in \mathbb{R}^{N \times N}$ that maps the \emph{ideal} computational-basis populations of the post-circuit state to the \emph{observed} outcome frequencies, that is,
\begin{equation}
    z = A x.
    \label{eq:Ax_classical}
\end{equation}
This picture underlies a wide range of readout calibration and error-mitigation techniques, as seen in \cite{Geller2020}, \cite{Jattana2020}, \cite{Hamilton2020}, \cite{Döbler2024}, \cite{Nation2021} and \cite{Gonzales2025}, among others. 

However, Eq.~\eqref{eq:Ax_classical} hides a strong physical assumption: it implicitly assumes that the effective measurement is described by POVM elements that are \emph{diagonal} in the computational basis. Under this assumption, measurement outcomes depend only on the diagonal entries of the ideal density matrix $\tilde{\rho}$, i.e., on classical populations, and are completely insensitive to quantum coherences. In realistic devices, the measurement process is preceded by nontrivial dynamics, including coherent couplings, residual rotations, and possibly correlated noise. There is no reason, in general, for the effective POVM to be diagonal.

In this work, we revisit the assignment matrix model from first principles. Starting from an arbitrary completely positive trace-preserving (CPTP) channel acting before a computational-basis measurement, we derive an exact expression for the observed probabilities in terms of the populations and coherences of the ideal post-circuit state. The result takes the form
\begin{equation}
    z = A x + C y ,
    \label{eq:AxCy}
\end{equation}
where, $y \in \mathbb{R}^{N(N-1)}$ is a vector of coherence parameters (which is formed from the off-diagonal elements of $\tilde{\rho}$) and $C$ encapsulates the action these coherences have on the state population. 

As would be expected, there is a natural link between the $A$ and $C$ matrices, and the superoperator matrix representation of the quantum error channel, denoted by $H \in \mathbb{C}^{4^{n} \times 4^{n}}$. Quantum process tomography (QPT) refers to the class of techniques used to reconstruct $H$ from measurement data. Therefore, the elements of $A$ and $C$ can be found via QPT. There has been much work in QPT, for example \cite{Linh2025}, \cite{Flammia2012}, \cite{AbuGhanem2025}, \cite{Torlai2023}. However, the objective in QPT is a matrix of size $N^{2} \times N^{2}$, while the matrices $A$ and $C$ have sizes $N \times N$ and $N \times N(N-1)$, respectively. As such, the computational burden of finding $A$ and $C$ is far smaller than that of finding $H$. This cost reduction, of course, comes at the cost of forming the underdetermined system \eqref{eq:AxCy}, at which point one must refer to the literature of compressed sensing to solve, e.g., \cite{Nagahara2023} or \cite{Hosny2023}.

The standard assignment matrix model is recovered if and only if $C = 0$; in that case the POVM elements are diagonal. A simple case when the POVM elements are diagonal is when the noise channel is completely Pauli. Such a scenario holds, e.g., on Pauli twirled circuits \cite{Geller2013}. 

The rest of this paper is organized as follows. In Section~\ref{sec:derivation} we derive Eq.~\eqref{eq:AxCy} in detail. Section~\ref{sec:Cmeaning} analyzes the structure and properties of $A$ and $C$. Section~\ref{sec:examples} illustrates the decomposition for several representative channels. Section~\ref{sec:implications} summarizes the implications for readout modeling and device characterization. Section~\ref{sec:numerical results} illustrates the potential advantages of the full $z = Ax + Cy$ model through numerical experiments on multi-qubit systems. Section~\ref{sec:conclusions} summarizes our findings and discusses directions for future work.

\section{General framework and derivation}
\label{sec:derivation}

\subsection{Setup}

We consider a generic circuit on $n$ qubits, represented by a unitary $U$ acting on an initial state $\rho$. We denote the Hilbert space spanned by the qubits as $\mathcal{H} \sim \mathbb{C}^{N}$, where $N = 2^n$. The ideal post-circuit state is
\begin{equation}
    \tilde{\rho} = U \rho U^\dagger.
\end{equation}
In practice, the state is subject to noise before measurement. We describe this noise by a completely positive, trace preserving (CPTP) channel
\begin{equation*}
    \mathcal{E} : \mathcal{H} \rightarrow \mathcal{L}(\mathcal{H}),
\end{equation*}
where $\mathcal{L}(\mathcal{H})$ denotes the space of linear operators on $\mathcal{H}$. See \cite[p. 361]{NielsenChuang2010} for a discussion on CPTP error channels. By Kraus' theorem \cite[Theorem 8.3]{NielsenChuang2010}, there exist operators $\{E_a\}$ (which are called Kraus operators) such that
\begin{equation} \label{Eq: Full error channel}
    \mathcal{E}(\cdot) = \sum_a E_a (\cdot) E_a^\dagger, \quad \sum_{a} E_{a} E_{a}^{\dagger} = \mathbb{1}.
\end{equation}
The actual, \textit{noisy}, state prior to measurement is
\begin{equation}
    \rho' = \mathcal{E}(\tilde{\rho}).
\end{equation}

We assume that the projective measurement at the hardware level is in the computational basis $\{\ket{k}\}_{k=0}^{N-1}$. The probability of observing outcome $k$ is then
\begin{equation}
    z_k = \Pr(\text{outcome }k) = \Tr(\ket{k}\bra{k}\rho').
\end{equation}
It is convenient to ``push'' the noise channel onto the measurement operators and work in the Heisenberg picture. Recall that a positive operator-valued measure on $\mathcal{H}$ is a collection of positive semi-definite operators $\{ P_{m} \} \subset \mathcal{L}(\mathcal{H})$ such that $\sum_{m} P_{m} = \mathbb{1}$ (see, e.g., \cite[Section 2.2.6]{NielsenChuang2010} for a high-level discussion on POVMs, the reader may also be interested in \cite{Farenick2011} or \cite{Cedeñopérez2025} for a more rigorous, probabilistic interpretation of POVMs). For example, the measures $\{P_{m} \}$ defined by $P_{m} := E_{m} E_{m}^{\dagger}$ form a POVM.

We shall define a POVM to help characterize the error channel, let
\begin{equation}
    F_k := \mathcal{E}^\dagger(\ket{k}\bra{k})
        = \sum_a E_a^\dagger \ket{k}\bra{k} E_a.
    \label{eq:Fk_def}
\end{equation}
Each $F_k$ is a positive semi-definite operator (as it is a sum of positive semi-definite operators), and the completeness relation
$\sum_k \ket{k}\bra{k} = \mathbb{1}$ together with trace preservation of $\mathcal{E}$ implies
\begin{equation}
    \sum_k F_k = \mathcal{E}^\dagger(\mathbb{1}) = \mathbb{1}.
\end{equation}
Hence, $\{F_k\}$ is a POVM. We can then write
\begin{equation}
    \begin{aligned}
        z_k &= \Tr(|k \rangle \langle k | \rho') \\
        &= \Tr\left(|k \rangle \langle k | \sum_{a} E_{a}^{\dagger} \tilde{\rho} E_{a} \right) \\
        &= \sum_{a} \Tr (|k \rangle \langle k | E_{a} \tilde{\rho}E_{a}) \\
        &= \sum_{a} \Tr (E_{a} |k \rangle \langle k | E_{a} \tilde{\rho})\\
        &=\Tr(F_k \tilde{\rho}),
    \end{aligned}
    \label{eq:zk_trace}
\end{equation}
where the third to fourth equality was obtained by the cyclic property of the trace operation. Indeed, we see that we may use $\{ F_{k} \}$ to define a discrete probability measure $\Tr (F_{\cdot} \tilde{\rho})$. Now, we shall expand both $F_k$ and $\tilde{\rho}$ in a convenient operator basis and obtain some more insight into the mapping $\tilde{\rho} \rightarrow \rho'$.

\subsection{Recovering the vector form} \label{subsec: decomposition of the state}

We fix the computational basis $\{\ket{\ell}\}$. By definition of $x$, we have $x_{l} = \langle \ell | \tilde{\rho} | \ell \rangle$. Let us denote the off-diagonal elements by
\begin{equation}
    \begin{aligned}
        c_{lr} := \langle \ell | \tilde{\rho} | r \rangle.
    \end{aligned}
\end{equation}
By hermiticity of $\tilde{\rho}$ (i.e., $\tilde{\rho}^{\dagger} = \tilde{\rho}$), we see that $c_{\ell r} = c_{r \ell}^{*}$. Let $c^{R}_{\ell r}, c^{I}_{\ell r}$ denote the real and imaginary parts of $c_{\ell r}$, respectively. We may accordingly expand the density matrix $\tilde{\rho}$ as
\begin{equation}
    \begin{aligned}
        \tilde{\rho} &= \sum_{\ell} x_\ell \ket{\ell}\bra{\ell} + \sum_{\ell<r} \Big[ c^{(R)}_{\ell r} (\ket{\ell}\bra{r} + \ket{r}\bra{\ell})\\
        &\quad 
    + c^{(I)}_{\ell r} i (\ket{\ell}\bra{r} - \ket{r}\bra{\ell}) \Big].
    \end{aligned}
    \label{eq:rho_decomp}
\end{equation}

The terms in Eq.~\eqref{eq:rho_decomp} are arranged so that all coefficients are real. We may view \eqref{eq:rho_decomp} as a decomposition of $\tilde{\rho}$ into the operator basis $\{ | \ell \rangle \langle r |\}_{\ell, r}$. A straightforward calculation shows that this basis is orthonormal with respect to the Hilbert-Schmidt inner product
\begin{equation*}
    \langle B , D \rangle_{HS} = \Tr(B^{\dagger} D). 
\end{equation*}

We expand each POVM element $F_k$ in the same operator basis:
\begin{equation}
    \begin{aligned}
        F_k &= \sum_{\ell} \alpha_{k\ell} \ket{\ell}\bra{\ell} + \sum_{\ell<r} \Big[ \beta^{(R)}_{k,\ell r} (\ket{\ell}\bra{r} + \ket{r}\bra{\ell}) \\
        &\quad 
        + \beta^{(I)}_{k,\ell r} i (\ket{\ell}\bra{r} - \ket{r}\bra{\ell}) \Big].
    \end{aligned}
    \label{eq:Fk_decomp}
\end{equation}
The real coefficients are given by
\begin{align}
    \alpha_{k\ell} &= \braket{\ell|F_k|\ell} \ge 0,
    \label{eq:alpha_def} \\
    \beta^{(R)}_{k,\ell r} &= \Re \braket{r|F_k|\ell}, \\
    \beta^{(I)}_{k,\ell r} &= \Im \braket{r|F_k|\ell}.
\end{align}
We let $\beta_{k, \ell r} = \langle r | F_{k} | \ell \rangle$, so $\beta_{k, \ell r} = \beta^{(R)}_{k \ell r} + i \beta^{(I)}_{k, \ell r}$. The positivity of $F_k$ ensures that these coefficients are not arbitrary, but we will not need their detailed constraints here.

The POVM completeness condition $\sum_k F_k = \mathbb{1}$ implies that
\begin{equation}
    \sum_k \alpha_{k\ell} = \sum_{k} \langle \ell | F_{k} | \ell \rangle = 1, \quad \forall \ell,
    \label{eq:alpha_colsums}
\end{equation}
and
\begin{equation}
    \sum_{k} \beta_{k, \ell r} = \sum_{k} \langle \ell | F_{k} | r \rangle = 0, \quad \forall \ell < r,
\end{equation}
which implies that
\begin{equation}
    \sum_k \beta^{(R)}_{k,\ell r} = 0, \quad
    \sum_k \beta^{(I)}_{k,\ell r} = 0, \quad \forall \ell < r.
\end{equation}

To compute $z_k = \text{Tr}(F_k \tilde{\rho})$, we can rely on the linearity of the trace and the orthogonality of the basis operators under the Hilbert-Schmidt inner product. A straightforward, but tedious computation given in Appendix~\ref{App: derivation of zk_master} shows that
\begin{equation} \label{eq:zk_master}
    z_k = \sum_{\ell} \alpha_{k\ell} x_\ell + 2 \sum_{\ell < r} \left( \beta^{(R)}_{k,\ell r} c^{(R)}_{\ell r} + \beta^{(I)}_{k,\ell r} c^{(I)}_{\ell r} \right).
\end{equation}
This is the central relation at the level of components. 

We now collect terms into a compact vector equation. Let $x \in \mathbb{R}^N$ be the population vector with entries $x_\ell$. We shall gather $y \in \mathbb{R}^{N (N-1)}$ in a standard lexicographical stacking of the $c_{\ell r}$ terms
\begin{equation}
    y := \begin{bmatrix}
        c_{0,1}^{(R)} & c_{0,1}^{(I)} & \cdots & c_{N-2,N-1}^{(R)} & c_{N-2, N-1}^{(I)}
    \end{bmatrix}^{T}.
\end{equation}
We define the $k$-th row of $C \in \mathbb{R}^{N \times N (N-1)}$ accordingly
\begin{equation}
    [C]_{k} := \begin{bmatrix}
        \beta_{k,0,1}^{(R)} & \beta_{k, 0,1}^{(I)} & \cdots & \beta_{k, N-2, N-1}^{(R)} & \beta_{k, N-2, N-1}^{(I)}
    \end{bmatrix}.
\end{equation}
Then, \eqref{eq:zk_master} is equivalent to
\begin{equation}
    z = A x + C y,
\end{equation}
which is exactly \eqref{eq:AxCy}.

\subsection{An alternative characterization}
We can also represent the quantum error channel via an \textit{operator-valued} kernel $K : \mathcal{H} \times \mathcal{H} \rightarrow \mathcal{L}(\mathcal{H})$ defined by
\begin{equation} \label{Def: Quantum kernel}
    K(s,t) := \sum_{a} E_{a} |s \rangle \langle t| E_{a}^{\dagger} = \mathcal{E}(|s\rangle \langle t|).
\end{equation}
Intuitively, the operator-valued kernel describes how the \textit{ideal} (albeit potentially non-physical) state $| \ell \rangle \langle r|$ is mapped under the noise channel. The hope is that we may eventually utilize the rich literature of operator-valued kernels, e.g., \cite{Jorgensen2025} and \cite{Jorgensen2025B} to bring more insight into our problem.

We first deduce an equivalent condition that results in the assignment matrix relationship \eqref{eq:Ax_classical}. By our decomposition
\begin{equation}
    \tilde{\rho} = \sum_{\ell} x_{\ell} | \ell \rangle \langle \ell | + \sum_{\ell \neq r} c_{\ell r} | \ell \rangle \langle r |
\end{equation}
introduced in section~\ref{subsec: decomposition of the state}, and starting with \eqref{eq:zk_trace}, we see that
\begin{equation}
    \begin{aligned}
        z &= \text{Tr}(F_{k} \tilde{\rho}_{k}) \\
        &= \text{Tr} \left( \sum_{\ell} x_{\ell} F_{k} | \ell \rangle \langle \ell| + \sum_{\ell \neq r} c_{\ell r} F_{k} |\ell \rangle \langle r| \right) \\
        &= \sum_{\ell} x_{\ell} \sum_{a} \text{Tr}(E_{a}^{\dagger} | k \rangle \langle k| E_{a} | \ell \rangle \langle \ell|) \\
        &\quad + \sum_{\ell \neq r} c_{\ell r} \sum_{a} \text{Tr}( E_{a}^{\dagger} |k \rangle \langle k | E_{a} | \ell \rangle \langle r|) \\
        &= \sum_{\ell} x_{\ell} \text{Tr}(K(\ell, \ell) | k \rangle \langle k |) \\
        &\quad + \sum_{\ell \neq r} c_{\ell r} \text{Tr}(K(\ell, r) | k \rangle \langle k |).
    \end{aligned}
\end{equation}
Then, we see that
\begin{equation} 
    z_{k} = \sum_{\ell} x_{\ell} \langle k | K(\ell,\ell) | k \rangle + \sum_{\ell \neq r} c_{\ell, r} \langle k | K(\ell,r) | k \rangle.
\end{equation}
In this case, we recover $Ax=z$ if and only if 
\begin{equation} \label{Eq: Iff Cond for Correction Matrix}
        \operatorname{Tr}[|k \rangle \langle k| K(\ell,r)] = 0, \; \text{for all} \;\ell \neq r
\end{equation}

\section{The coherence-response matrix and its physical meaning}
\label{sec:Cmeaning}

The new ingredient in Eq.~\eqref{eq:AxCy} is the matrix $C$, which we now discuss in more detail. By construction, $C$ is defined so that
\begin{equation}
    (C y)_k =
    \sum_{\ell<r} \bigg[
        2 \beta^{(R)}_{k,\ell r} c^{(R)}_{\ell r}
        - 2 \beta^{(I)}_{k,\ell r} c^{(I)}_{\ell r}
    \bigg].
\end{equation}
In other words, each row of $C$ contains the coefficients $\beta^{(R)}_{k,\ell r}$ and $\beta^{(I)}_{k,\ell r}$ that specify how the coherences of the state enter into the probability of outcome $k$.

It follows immediately that $C = 0$ if and only if
\begin{equation}
    \beta^{(R)}_{k,\ell r} = \beta^{(I)}_{k,\ell r} = 0
\end{equation}
for all $k$ and all $\ell < r$. In turn, this is equivalent to requiring that all POVM elements $F_k$ are diagonal in the computational basis:
\begin{equation}
    \braket{r|F_k|\ell} = 0, \quad \forall k, \forall \ell \ne r.
\end{equation}
Thus the condition $C = 0$ is \emph{precisely} the condition that the measurement be insensitive to coherences of $\tilde{\rho}$ in the computational basis. By \eqref{Eq: Iff Cond for Correction Matrix}, we can see that asking for \textit{diagonal} POVM elements is equivalent to asking for operator-valued kernels $K(\ell, r)$ which have \textit{zero} diagonal for any $\ell \neq r$.

The matrix $A$ defined above coincides with the usual assignment matrix familiar from readout calibration. When $C = 0$, its entries have a simple interpretation: $A_{k\ell}$ is the probability of obtaining outcome $k$ when the system is prepared in the computational basis state $\ket{\ell}$. When $C \neq 0$, then $A$ quantifies the `classical' effects, rather than the coherences. 

\subsection{Dependence on coherences}

Whenever $C \neq 0$, the measured probabilities $z_k$ depend on the off-diagonal elements of $\tilde{\rho}$. This dependence is linear in the real and imaginary parts $c^{(R)}_{\ell r}$ and $c^{(I)}_{\ell r}$, with coefficients determined by the off-diagonal matrix elements of $F_k$.

Physically, this means that the measurement is sensitive to the \emph{phase} information encoded in superpositions of computational-basis states. For a given pair $(\ell,r)$, the relative weight of $c^{(R)}_{\ell r}$ and $c^{(I)}_{\ell r}$ in $(C y)_k$ indicates whether the probability of outcome $k$ depends more strongly on cosine-like or sine-like components of the relative phase.

\subsection{Interpretation as a ``non-classicality'' measure}

From the perspective of readout modeling, $A$ captures the purely classical confusion between basis states, while $C$ captures all coherent contributions. One can therefore regard $C$ (or appropriate norms of $C$) as a quantitative measure of how far the effective measurement is from an ideal classical assignment. In the extreme case $C = 0$, the measurement is entirely classical in the computational basis; all deviations from that scenario are encoded in $C$.

In multi-qubit devices, the pattern of nonzero entries in $C$ can also reveal structure: which pairs of basis states interfere in which outcomes, whether the measurement couples primarily states that differ by single-qubit flips or more complex patterns, and whether there are signatures of correlated or entangling dynamics in the readout.

\section{Examples}
\label{sec:examples}

We now illustrate the decomposition for some representative noise channels.

\subsection{Pure dephasing}

Consider pure dephasing in the computational basis. For a single qubit, the channel can be written as
\begin{equation}
    \mathcal{E}(\rho) =
    \begin{pmatrix}
        \rho_{00} & \lambda \rho_{01} \\
        \lambda \rho_{10} & \rho_{11}
    \end{pmatrix},
\end{equation}
with $|\lambda| \le 1$. In the Heisenberg picture, the POVM elements are unchanged:
\begin{equation}
    F_0 = \ket{0}\bra{0},\qquad F_1 = \ket{1}\bra{1}.
\end{equation}
These are diagonal in the computational basis, so $\beta^{(R)}_{k,\ell r} = \beta^{(I)}_{k,\ell r} = 0$ and $C = 0$. The assignment matrix is $A = \mathbb{1}$. The measurement is insensitive to coherences, as expected for pure dephasing in the measurement basis.

\subsection{Amplitude damping}

For single-qubit amplitude damping with parameter $\gamma \in [0,1]$, an operator-sum representation is
\begin{equation}
    E_0 =
    \ket{0}\bra{0} + \sqrt{1-\gamma}\ket{1}\bra{1}, \qquad
    E_1 = \sqrt{\gamma} \ket{0}\bra{1}.
\end{equation}
One finds
\begin{align}
    F_0 &= \mathcal{E}^\dagger(\ket{0}\bra{0})
    = \ket{0}\bra{0} + \gamma \ket{1}\bra{1}, \\
    F_1 &= \mathcal{E}^\dagger(\ket{1}\bra{1})
    = (1-\gamma) \ket{1}\bra{1}.
\end{align}
Both $F_0$ and $F_1$ are diagonal. Therefore $C = 0$, and
\begin{equation}
    A = 
    \begin{pmatrix}
        1 & \gamma \\
        0 & 1-\gamma
    \end{pmatrix}.
\end{equation}
Even though amplitude damping is non-Pauli and changes populations, it leaves the POVM diagonal in the computational basis, so measurement remains classical in this sense.

\subsection{Coherent over-rotation before measurement}

Now consider a coherent unitary rotation immediately before measurement. For a single qubit, let
\begin{equation}
    U = R_y(\theta) = e^{-i \theta Y / 2},
\end{equation}
and define a noise channel $\mathcal{E}(\rho) = U \rho U^\dagger$. Then
\begin{equation}
    F_k = U^\dagger \ket{k}\bra{k} U.
\end{equation}
An explicit calculation yields
\begin{align}
    F_0 &= 
    \begin{pmatrix}
        \cos^2(\theta/2) & \tfrac12 \sin\theta \\
        \tfrac12 \sin\theta & \sin^2(\theta/2)
    \end{pmatrix},
    \\
    F_1 &=
    \begin{pmatrix}
        \sin^2(\theta/2) & -\tfrac12 \sin\theta \\
        -\tfrac12 \sin\theta & \cos^2(\theta/2)
    \end{pmatrix}.
\end{align}
The off-diagonal entries are nonzero for $\theta \ne 0$. For a general state
\begin{equation}
    \tilde{\rho} =
    \begin{pmatrix}
        x_0 & c_{01} \\
        c_{01}^* & x_1
    \end{pmatrix},
\end{equation}
with $c_{01} = c^{(R)}_{01} + i c^{(I)}_{01}$, the probabilities become
\begin{align}
    z_0 &= \cos^2(\tfrac{\theta}{2})x_0 + \sin^2(\tfrac{\theta}{2})x_1 + \sin\theta\, c^{(R)}_{01},
    \\
    z_1 &= \sin^2(\tfrac{\theta}{2})x_0 + \cos^2(\tfrac{\theta}{2})x_1 - \sin\theta\, c^{(R)}_{01}.
\end{align}
Comparing with Eq.~\eqref{eq:AxCy}, we see that
\begin{equation}
    A =
    \begin{pmatrix}
        \cos^2(\theta/2) & \sin^2(\theta/2) \\
        \sin^2(\theta/2) & \cos^2(\theta/2)
    \end{pmatrix},
\end{equation}
and 
\begin{equation}
    C = \begin{pmatrix}
        \sin \theta & 0 \\ -\sin \theta & 0
    \end{pmatrix}.
\end{equation}
Thus, $C \neq 0$ whenever $\theta \ne 0$.

Physically, this example shows that even a simple coherent misalignment of the measurement axis leads to $C \neq 0$: the measurement probabilities become explicitly sensitive to the real part of the coherence between $\ket{0}$ and $\ket{1}$. If one were to ignore $C$ and attempt to model this situation with a classical assignment matrix alone, the resulting model would necessarily be incomplete, because it would predict $z$ as a function of $x$ only and would miss the dependence on $c^{(R)}_{01}$.

\section{Implications for readout modeling}
\label{sec:implications}

We have shown that, in full generality, the measurement statistics under arbitrary CPTP noise preceding a computational-basis measurement obey the linear relation 
\begin{equation}
    z = A x + C y,
\end{equation}
where $A$ is a column-stochastic assignment matrix acting on the populations $x$, and $C$ is a coherence-response matrix acting on the vector of coherences $y$. The classical model $z = A x$ corresponds to the special case $C = 0$, i.e., to effective POVM elements that are diagonal in the computational basis.

From a practical perspective, this decomposition highlights a fundamental limitation of models that retain only $A$. Such models can accurately describe pure classical confusion and a subset of noise channels (such as pure dephasing and amplitude damping in the measurement basis), but they cannot capture readout effects that arise from coherent dynamics in the measurement chain. Any pre-measurement unitary rotation, coherent cross-talk, or entangling interaction that generates off-diagonal entries in $F_k$ will necessarily give $C \neq 0$ and hence introduce coherence sensitivity into the measurement.

The matrix $C$ itself contains rich information about the device. Its entries encode which pairs of computational-basis states interfere in which measurement outcomes, with what signs and magnitudes, and whether the measurement is more sensitive to the real or imaginary parts of specific coherences. In multi-qubit settings, patterns in $C$ may reveal locality structure, correlated errors, or signatures of nontrivial Hamiltonian dynamics during readout.

These observations suggest that a more faithful description of real-device readout should, when appropriate, include both $A$ and $C$. In the rest of this section, we shall discuss the computational complexity of obtaining these matrices, solving \eqref{eq:AxCy}, and their comparison to QPT.

\subsection{Forming the Systems}
In this subsection, we compare the cost of obtaining the elements in $A$ and $C$ as compared to the full superoperator matrix, i.e., the matrix which solves
\begin{equation} \label{eq:superoperator}
    \text{vec}(\rho') = H \text{vec}(\tilde{\rho}),
\end{equation}
where $\text{vec}(D)$ denotes the typical column-stacked vectorization of matrix $D$. In order to find $\tilde{\rho}$, we need the noisy density matrix $\rho'$ and the superoperator matrix $H$. To find the ideal output distribution $x$, we will need the noisy distribution $z$ and the matrices $A$ and $C$. We shall compare the cost of obtaining these relevant ingredients.

First, let us compare the cost of obtaining $\rho' \in \mathbb{C}^{N \times N} $ against $z \in \mathbb{R}^{N}$. The process of obtaining the former is called \textit{quantum state tomography}. This has been extensively studied in the literature, and we briefly review their techniques and costs. For a system of $n$ qubits and $n$ detectors, one can directly measure $\rho'$ using at least $4^{n}-1$ total measurements, as described in \cite[Section 2]{Altepeter2004}. Beyond the simple measurement techniques outlined in \cite{Altepeter2004}, there is a whole plethora of methods available to estimating $\rho'$, each with their own advantages and disadvantages. Some of these methods use linear inversion \cite{Vogel1989}, \cite[Section 2.2.1]{Nielsen2021}, maximum likelihood estimation \cite{Rehacek2001}, least squares estimation \cite{Lum2022}, \cite{Goncalves2011}, and even some machine-learning based approaches \cite{Innan2024}, \cite{Wu2025}. Regardless of the method we use, one must still obtain $N^{2}$ elements to find $\rho'$. In contrast, $z$ has only $N$ entries and can be easily estimated via Monte-Carlo simulations and solely measuring in the computational basis.

Next, let us compare the matrices involved. $H$ is an $N^{2} \times N^{2}$ matrix while $A$ is $N \times N$ and $C$ is $N \times N(N-1)$. The act of finding $H$ (or equivalently, finding the Kraus operators $E_{a}$ in \eqref{Eq: Full error channel}) is called \textit{quantum process tomography} (QPT). This process involves preparing multiple different ideal states $\tilde{\rho}_{j}$, using state tomography to find the associated noisy states $\rho'_{j}$, and then assembling these together to infer the noise channel $\mathcal{E}$ (and therefore find $H$). This technique is outlined in, e.g., \cite{Leung2002}, \cite[Section 3.2]{Greenbaum2015}, \cite[Section 2.2.2]{Nielsen2021}. Regardless, $N^{4} = 16^{n}$ elements are required. The matrices $A$ and $C$ have $N^{3}=8^{n}$ elements to be determined. In other words, finding the $H$ matrix is a problem that is $2^{n}$ times bigger than finding $A$ and $C$. 

\subsection{Solving the Systems}
Once the matrices $\rho'$ and $H$ are determined, there is still the problem of solving \eqref{eq:superoperator}. Generally, one can resort to any number of popular diagonalization techniques or use the pseudoinverse to solve this problem \cite[Section 2]{Nielsen2021}. More sophisticated techniques, however, do exist, such as the method proposed in \cite{Cattaneo2023}. 

To solve \eqref{eq:zk_master}, we shall define
\begin{equation*}
    B := \begin{bmatrix}
        A & C
    \end{bmatrix}, \quad v := \begin{bmatrix}
        x \\ y
    \end{bmatrix}.
\end{equation*}
In this case, we have
\begin{equation} \label{Eq: What we actually need to solve}
    z = Bv.
\end{equation}
While the system \eqref{Eq: What we actually need to solve} has far smaller dimensionality than \eqref{eq:superoperator}, this comes at the trade-off of being vastly underdetermined (note that $B$ has size $N \times N^{2}$). This means that \eqref{Eq: What we actually need to solve} will typically have infinitely many solutions. However, we know that the desired solution $v^{*}$ to \eqref{Eq: What we actually need to solve} must be physically viable. Let $\mathcal{P} \subset \mathbb{R}^{N^{2}}$ denote the set of physically acceptable vectors. Then, we can reformulate solving \eqref{Eq: What we actually need to solve} as solving
\begin{equation} \label{More tangible problem}
    \inf \left\{ ||z-Bv|| : v \in \mathcal{P} \right\},
\end{equation}
where $||\cdot||$ is a suitable norm.

We shall leave the question of solving \eqref{Eq: What we actually need to solve} and \eqref{More tangible problem} to future work. However, the question of solving underdetermined systems has been extensively studied in the literature (e.g., \cite{Kuegler2012}, \cite{Cline1976}, \cite{Donoho2006}, \cite{Hosny2023}, and \cite{Nagahara2023}, among others), so solving \eqref{More tangible problem} is not an entirely novel, untested, idea. Indeed, variations of this problem include the well-studied least-squares problem (as $||\cdot||$ becomes the Euclidean norm), which has been used in countless real-world problems in the past several decades \cite{Lines1984}, \cite{Opfell1958}, \cite{Galindo-Prieto2026}, \cite{Miller2025}. 

\section{Numerical results}
\label{sec:numerical results}

This section evaluates potential performance gains from using the coherence-aware readout model over classical or model-free approaches. For our evaluations, we quantify the performance using the Hellinger Fidelity, 

\begin{equation}
    F_H(p,q) = \left(\sum_x \sqrt{p(x)\, q(x)}\right)^2,
\end{equation} 

where $p(x)$ is the ideal readout probability of bit string $x$, and $q(x)$ is the noisy readout probability of $x$. For the purposes of this work, we do not consider shot noise and simply take the exact probability distributions computed analytically from the measurement channel model.

\begin{figure}[htb]
    \centering
    \includegraphics[width=\columnwidth]{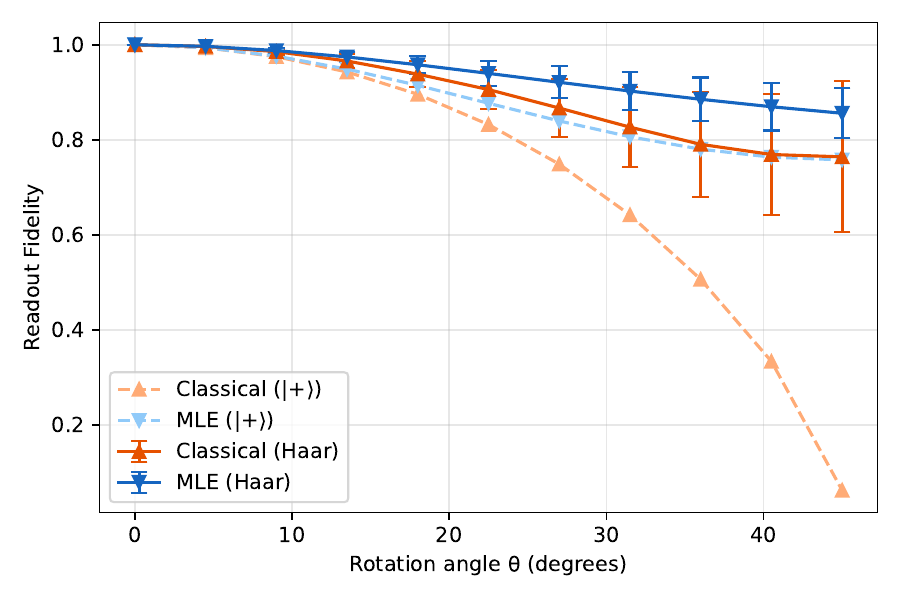}
    \caption{Readout fidelity vs.\ coherent over-rotation angle $\theta$ for a 4-qubit system with amplitude damping $\gamma = 0.01$, comparing MLE recovery (accounting for the full $z = Ax + Cy$ model) against the classical baseline ($A^{-1}z$). Results are shown for two input ensembles: the maximally coherent state $|{+}\rangle^{\otimes 4}$ (dashed) and 100 Haar-random pure states (solid, with error bars indicating $\pm 1$ standard deviation across states).}
    \label{fig:mle_vs_classical}
\end{figure}

Fig.~\ref{fig:mle_vs_classical} compares the resulting readout fidelities of the classical model to our coherence-aware model on a $4$-qubit circuit with $R_y$ over-rotation applied to each qubit before measurement. Each angle is evaluated over 100 Haar-random pure states as well as the all-plus state $|{+}\rangle^{\otimes n}$. To solve the under-determined system (Eq.~\ref{Eq: What we actually need to solve}), we use the Maximum-likelihood estimator (MLE), formulated as a convex optimization problem over the space of valid density matrices, 

\begin{align*}
    \hat{\rho} &= \underset{\rho}{\arg\max} \sum_{k} z_k \log p_k(\rho) \label{eq:mle} \\
    &\text{subject to} \quad \rho \succeq 0, \quad \operatorname{Tr}(\rho) = 1 \nonumber, 
\end{align*}  

where $z_k$ are the observed outcome frequencies and $p_k(\rho) = (Ax + Cy)_k$ is the predicted probability for outcome $k$ under our linear model. This objective is equivalent to minimising the KL divergence $\mathrm{KL}(z |p(\rho))$. The classical baseline we compare to applies the standard calibration matrix inversion $\hat{x} = A^{-1}z$. As can be seen in Fig.~\ref{fig:mle_vs_classical}, as $\theta$ increases, the mean readout fidelity drops compared to the coherence-aware MLE approach, and the variance across states increases. Notice that the classical approach performs significantly worse for the all-plus state $|{+}\rangle^{\otimes n}$, where the coherence contribution $Cy$ is maximized. 

\begin{figure}[htb]
    \centering
    \includegraphics[width=\columnwidth]{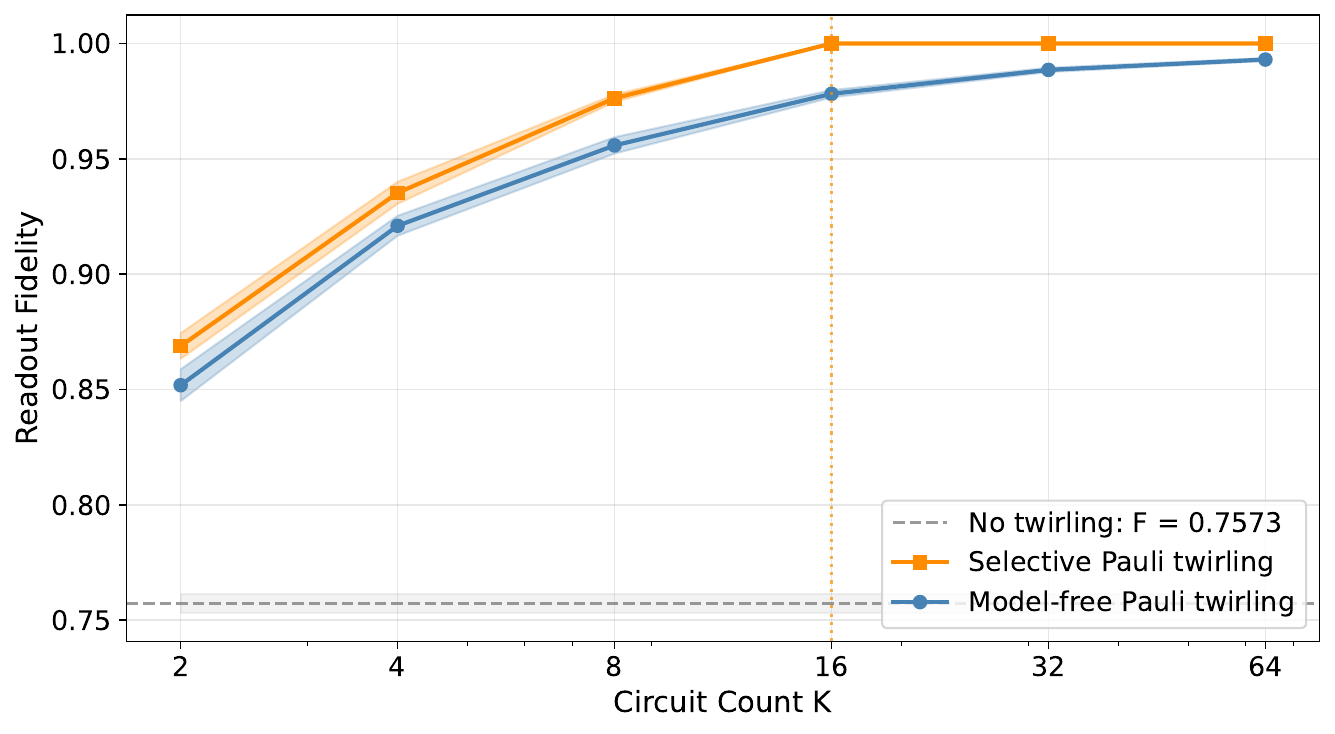}
    \caption{Readout fidelity vs.\ circuit count $K$ for a 6-qubit system with coherent over-rotation ($\theta = 45°$) on 2 qubits and no noise on the remaining 4, averaged over 100 Haar-random pure states (shaded bands indicate $\pm 1$ standard deviation). Selective Pauli twirling applies the balanced support condition over only the $\tau = 2$ noisy qubits, requiring $4^\tau = 16$ circuits to saturate at perfect fidelity. Model-free Pauli twirling randomises all $n = 6$ qubits uniformly and has not converged by $K = 64$, requiring $4^n = 4096$ circuits in principle. The dotted vertical line marks $K = 4^\tau = 16$, the saturation threshold for selective twirling. The dashed horizontal line shows the no-twirling baseline ($F = 0.757$), where unmitigated coherent readout noise suppresses population recovery.}
    \label{fig:sel_pauli_advantage}
\end{figure}

The coherence-response matrix can also be leveraged to improve the sampling efficiency of Pauli twirling. In model-free Pauli twirling, the expected number of random circuits required to cancel the effects of coherent readout error is on the order of $4^n$, where $n$ is the number of qubits. However, if coherent noise is concentrated on only a subset of $m \leq n$ qubits, then the structure of $C$ can be leveraged to identify these qubits and tailor Pauli-twirling sampling for faster convergence. Assuming independent per-qubit noise, the POVM elements factorize as $F_x = \bigotimes_{i=1}^n F_{x_i}^{(i)}$, which gives a natural partitioning of the columns of $C$ by which qubit's off-diagonal response they probe. Specifically, we define the per-qubit coherence submatrix $C^{(i)}$ as the restriction of $C$ to columns corresponding to coherence pairs $(l, r)$ that differ only in bit position $i$, and compute the per-qubit diagnostic score

$$\Gamma_i = |C^{(i)}|_F,$$

where $\|\cdot\|_F$ is the Frobenius norm. A large $\Gamma_i$ indicates that qubit $i$ is a dominant source of coherent readout noise. For an Ry over-rotation of angle $\theta_i$ one finds $\Gamma_i \propto |\sin\theta_i|$, while $\Gamma_i = 0$ for any qubit with a diagonal POVM element.

In analogy with the subsystem-balanced Pauli twirling approach of \cite{xu2025efficientmeasurementerrormitigation}, a balanced sample over the $4^m$ Pauli combinations on the support qubits can be drawn during twirling to guarantee that every element of the support subgroup is sampled equally. This ensures exact cancellation of the coherent noise contribution from the $m$ dominant qubits after $4^m$ circuit samples. Model-free twirling, in contrast, achieves cancellation only in expectation and requires $O(4^n)$ random circuits. 

As an example, for a 6-qubit circuit in which 2 qubits have dominant coherent Ry rotation error at degree $\theta$, only $4^2 = 16$ circuits are required to exactly cancel the coherent readout noise from the $2$ qubits. Compared to model-free twirling over all $6$ qubits, this leads to a $256\times$ reduction in circuit overhead. This improved efficiency is illustrated in Fig.~\ref{fig:sel_pauli_advantage}. It can be seen that the $C$-matrix-guided balanced approach reaches its fidelity ceiling of $1.00$ at exactly $K = 16$ Pauli-twirled circuits, at which point the coherent noise from the two dominant qubits is exactly canceled. Model-free twirling approaches the same ceiling only asymptotically.

\section{Conclusions}
\label{sec:conclusions}

We have given a detailed derivation and analysis of the decomposition
\begin{equation}
    z = A x + C y
\end{equation}
for computational-basis measurements under arbitrary CPTP noise. The familiar assignment matrix $A$ captures the effects of classical confusion between basis states, while the coherence-response matrix $C$ captures the contributions arising from off-diagonal components of the effective POVM. The condition $C = 0$ is equivalent to asking that the POVM $\{F_{k} \}$ defined by \eqref{eq:Fk_def} is diagonal in the computational basis. As such, the widely used classical readout model $z = A x$ (as used in \cite{Nation2021}, \cite{Gonzales2025}, \cite{Geller2020}, \cite{Bravyi2021}, among others) is valid if and only if the measurement is entirely insensitive to coherences of the state in that basis.

By working through explicit examples, we have shown that $C$ vanishes for pure dephasing and amplitude damping in the measurement basis, but is nonzero for coherent over-rotations, illustrating how even simple coherent distortions generate coherence-sensitive readout. Indeed, a Pauli-twirled circuit (i.e., the process described in \cite{Geller2013}) ideally has a purely Pauli noise channel, and therefore the associated POVM is diagonal in the computational basis. In general, however, we show that $C$ contains valuable information about coherent aspects of the measurement process---including hidden basis misalignment and interference between computational-basis states---that is completely invisible in models that retain only $A$. 

Of course, one may use the full Liouville superoperator representation of the noise channel \eqref{eq:superoperator}, rather than \eqref{eq:AxCy}, to solve for the ideal state $\tilde{\rho}$. This effectively amounts to resorting to QPT for error characterization, which has been extensively explored; e.g., \cite{AbuGhanem2025}, \cite{Greenbaum2015}, \cite{Altepeter2004}, \cite{Cattaneo2023}, \cite{Flammia2012}, among others. However, our system \eqref{eq:AxCy} is of dimensionality $N \times N^{2}$, rather than $N^{2} \times N^{2}$. As $N = 2^{n}$, our system is thus exponentially smaller than \eqref{eq:superoperator}, and therefore will be far cheaper to solve.

Our numerical experiments reinforce these observations. On a 4-qubit system with coherent over-rotation, maximum-likelihood recovery using the full $z = Ax + Cy$ model maintains high fidelity at rotation angles where classical matrix inversion degrades, with the gap most pronounced for highly coherent input states such as $\ket{+}^{\otimes n}$. Moreover, the per-qubit diagnostic scores $\Gamma_i$ extracted from $C$ can be used to identify the dominant sources of coherent noise, enabling selective Pauli twirling that achieves exact cancellation with $4^m$ circuits on the $m$ noisy qubits rather than $O(4^n)$ for model-free twirling. The MLE approach, however, comes at the trade-off of working with a vastly underdetermined system. For future work, we propose to focus on solving a problem of the form \eqref{More tangible problem}, wherein we minimize a suitable norm while maintaining the physics of the problem. Both the classical results in compressed sensing, e.g., those reviewed in \cite{Hosny2023} and \cite{Nagahara2023}, as well as recent results with machine-learning techniques, such as \cite{Hyun2021} and \cite{Bertalan2025}, have been successful in solving underdetermined systems, so we expect to find similar success with our problem too.

% This, of course, comes at the trade-off of working with a vastly underdetermined system. For future work, we propose to focus on solving a problem of the form \eqref{More tangible problem}, wherein we minimize a suitable norm while maintaining the physics of the problem. Both the classical results in compressed sensing e.g., those reviewed in \cite{Hosny2023} and \cite{Nagahara2023}, as well as recent results with machine-learning techniques, such as \cite{Hyun2021} and \cite{Bertalan2025}, have been successful in solving underdetermined systems, so we expect to find similar success with our problem too.

This framework provides a conceptually simple and mathematically complete foundation for coherence-sensitive readout modeling on quantum devices. We invite future work on the experimental estimation of $C$, on solving \eqref{Eq: What we actually need to solve} and \eqref{More tangible problem}, and on the incorporation of coherence sensitivity into device characterization and error mitigation protocols.

\section{Acknowledgements}
We would like to thank Kaitlin Smith, Siddharth Vijaymurugan, and Alvin Gonzales for helpful discussions while preparing this work. ZM and ZHS
also acknowledge support by the U.S. Department of Energy
(DOE) under Contract No. DE-AC02-06CH11357, through the
Office of Science, Office of Advanced Scientific Computing
Research (ASCR) Exploratory Research for Extreme-Scale
Science and Accelerated Research in Quantum Computing.

\appendix
\section{Derivation of \eqref{eq:zk_master}} \label{App: derivation of zk_master}
By \eqref{eq:Fk_decomp} and \eqref{eq:rho_decomp}, we have that
\begin{equation}\label{Eq: Intermediate zk}
    \begin{aligned}
        \Tr (F_{k} \tilde{\rho}) &= \Tr \Bigg( \Big(\sum_{\ell} \alpha_{k\ell} \ket{\ell}\bra{\ell} + \sum_{\ell<r} \Big[ \beta^{(R)}_{k,\ell r} (\ket{\ell}\bra{r} +  \\
        &\quad + \ket{r}\bra{\ell}) + \beta^{(I)}_{k,\ell r} i (\ket{\ell}\bra{r} - \ket{r}\bra{\ell}) \Big] \Big) \\
        &\quad \Big( \sum_{\ell} x_\ell \ket{\ell}\bra{\ell} + \sum_{\ell<r} \Big[ c^{(R)}_{\ell r} (\ket{\ell}\bra{r} + \ket{r}\bra{\ell})\\
        &\quad 
    + c^{(I)}_{\ell r} i (\ket{\ell}\bra{r} - \ket{r}\bra{\ell}) \Big] \Big)\Bigg) \\
    &= \Tr \Bigg[ \Bigg( \sum_{\ell} \alpha_{k\ell} \ket{\ell}\bra{\ell} \Bigg) \Bigg( \sum_{\ell} x_{\ell} \ket{\ell}\bra{\ell} \Bigg) \Bigg] \\
    &\quad + \text{Tr}\Bigg[ \Bigg( \sum_{\ell<r} \beta^{(R)}_{k,\ell r} (\ket{\ell}\bra{r} + \ket{r}\bra{\ell}) \Bigg) \\
    &\quad \left( \sum_{\ell<r} c^{(R)}_{\ell r} (\ket{\ell}\bra{r} + \ket{\ell}\bra{r}) \right) \Bigg] \\
    &\quad + \text{Tr}\Bigg[ \Bigg( \sum_{\ell<r} \beta^{(I)}_{k,\ell r} i (\ket{\ell}\bra{r} - \ket{r}\bra{\ell}) \Bigg) \\
    &\quad \Bigg( \sum_{\ell<r} c^{(I)}_{\ell r} i (\ket{\ell}\bra{r} - \ket{\ell}\bra{r}) \Bigg) \Bigg].
    \end{aligned}
\end{equation}
The second equality comes from the othornomality of the operator basis, which implies that
\begin{equation}
    \begin{aligned}
        \Tr &\Bigg( \Big(\sum_{\ell} \alpha_{k\ell} \ket{\ell}\bra{\ell} \Big) \Big ( \sum_{\ell<r} \Big[ c^{(R)}_{\ell r} (\ket{\ell}\bra{r} + \ket{r}\bra{\ell})\\
        &\quad + c^{(I)}_{\ell r} i (\ket{\ell}\bra{r} - \ket{r}\bra{\ell}) \Big] \Big)\Bigg) = 0
    \end{aligned} 
\end{equation}
and
\begin{equation}
    \begin{aligned}
        \Tr &\Bigg( \Big( \sum_{\ell} x_\ell \ket{\ell}\bra{\ell} \Big) \Big( \sum_{\ell<r} \Big[ \beta^{(R)}_{k,\ell r} (\ket{\ell}\bra{r} + \ket{r}\bra{\ell})  \\
        &\quad + \beta^{(I)}_{k,\ell r} i (\ket{\ell}\bra{r} - \ket{r}\bra{\ell}) \Big]  \Big) \Bigg) = 0
    \end{aligned}
\end{equation}
as we must have $\Tr (|a \rangle \langle b|) = \langle b | a \rangle = 0$, for any basis elements $|a \rangle, |b \rangle$.

Next, we simplify the terms in \eqref{Eq: Intermediate zk}. First,
\begin{equation}
    \begin{aligned}
        \Tr \Bigg[ \Bigg( \sum_{\ell} \alpha_{k\ell} & \ket{\ell}\bra{\ell} \Bigg) \Bigg( \sum_{m} x_m \ket{m}\bra{m} \Bigg) \Bigg]  \\
        &=\sum_{\ell, m} \alpha_{k\ell} x_m \Tr(\ket{\ell}\bra{\ell}\ket{m}\bra{m}) \\
        &= \sum_{\ell, m} \alpha_{k\ell} x_m \delta_{\ell m}.
    \end{aligned}
\end{equation}
Next, we compute
\begin{equation}
    \begin{aligned}
        \text{Tr}\Bigg[ &\left( \sum_{\ell<r} \beta^{(R)}_{k,\ell r} (\ket{\ell}\bra{r} + \ket{r}\bra{\ell}) \right) \left( \sum_{p<q} c^{(R)}_{pq} (\ket{p}\bra{q} + \ket{q}\bra{p}) \right) \Bigg] \\
        &= \sum_{\ell<r} \beta^{(R)}_{k,\ell r} c^{(R)}_{\ell r} \text{Tr}\left[ (\ket{\ell}\bra{r} + \ket{r}\bra{\ell})^2 \right] \\
        &= 2  \sum_{\ell<r} \beta^{(R)}_{k,\ell r} c^{(R)}_{\ell r},
    \end{aligned}
\end{equation}
where, as before, the cross-terms vanish by orthonormality. Likewise, we compute
\begin{equation}
    \begin{aligned}
        \text{Tr}\Bigg[ &\Bigg( \sum_{\ell<r} \beta^{(I)}_{k,\ell r} i (\ket{\ell}\bra{r} - \ket{r}\bra{\ell}) \Bigg) \Bigg( \sum_{p<q} c^{(I)}_{pq} i (\ket{p}\bra{q} - \ket{q}\bra{p}) \Bigg) \Bigg] \\
        &= \sum_{\ell<r} \beta^{(I)}_{k,\ell r} c^{(I)}_{\ell r} (i^2) \text{Tr}\left[ (\ket{\ell}\bra{r} - \ket{r}\bra{\ell})^2 \right] \\
        &=  2 \sum_{\ell<r} \beta^{(I)}_{k,\ell r} c^{(I)}_{\ell r}.
    \end{aligned}
\end{equation}
Putting this all together yields \eqref{eq:zk_master}.

\bibliographystyle{siam}
\bibliography{bib}

@article {Jattana2020,
    AUTHOR = {Jattana, Manpreet Singh and Jin, Fengping and De Raedt, Hans
              and Michielsen, Kristel},
     TITLE = {General error mitigation for quantum circuits},
   JOURNAL = {Quantum Inf. Process.},
  FJOURNAL = {Quantum Information Processing},
    VOLUME = {19},
      YEAR = {2020},
    NUMBER = {11},
     PAGES = {Paper No. 414, 17},
      ISSN = {1570-0755,1573-1332},
   MRCLASS = {81P68},
  MRNUMBER = {4174678},
       DOI = {10.1007/s11128-020-02913-0},
       URL = {https://doi.org/10.1007/s11128-020-02913-0},
}

@article{Geller2020,
    author = {Geller, Michael},
    title = {Rigorous Measurement Error Correction},
    journal = {Quantum Science and Technology},
    volume = {5},
    year = {2020}
}

@article{Nation2021,
  title = {Scalable Mitigation of Measurement Errors on Quantum Computers},
  author = {Nation, Paul D. and Kang, Hwajung and Sundaresan, Neereja and Gambetta, Jay M.},
  journal = {PRX Quantum},
  volume = {2},
  issue = {4},
  pages = {040326},
  numpages = {9},
  year = {2021},
  month = {Nov},
  publisher = {American Physical Society},
  doi = {10.1103/PRXQuantum.2.040326},
  url = {https://link.aps.org/doi/10.1103/PRXQuantum.2.040326}
}

@misc{Gonzales2025,
      title={Quantum Error Correction Without Encoding via the Circulant Structure of Pauli Noise and the Fast Fourier Transform}, 
      author={Alvin Gonzales},
      year={2025},
      eprint={2501.01953},
      archivePrefix={arXiv},
      primaryClass={quant-ph},
      url={https://arxiv.org/abs/2501.01953}, 
}

@article{Chiu2025,
    author = {Chiu, Neng-Chun and Trapp, Elias C. and Guo, Jinen and Abobeih, Mohamed H. and Stewart, Luke M. and Hollerith, Simon and Stroganov, Pavel L. and Kalinowski, Marcin and Geim, Alexandra A. and Evered, Simon J. and Li, Sophie H. and Lyu, Xingjian and Peters, Lisa M. and Bluvstein, Dolev and Wang, Tout T. and Greiner, Markus and Vuleti\'{c} and Lukin, Mikhail D.},
    title = {Continuous operation of a coheren 3000-qubit system},
    journal = {Nature},
    volume = {646},
    issue = {8087},
    year = {2025}
}

@article{Bravyi2021,
  title = {Mitigating measurement errors in multiqubit experiments},
  author = {Bravyi, Sergey and Sheldon, Sarah and Kandala, Abhinav and Mckay, David C. and Gambetta, Jay M.},
  journal = {Phys. Rev. A},
  volume = {103},
  issue = {4},
  pages = {042605},
  numpages = {12},
  year = {2021},
  month = {Apr},
  publisher = {American Physical Society},
  doi = {10.1103/PhysRevA.103.042605},
  url = {https://link.aps.org/doi/10.1103/PhysRevA.103.042605}
}

@article{Guglielmo2024,
    author = {Mazzola, Guglielmo},
    title = {Quantum computing for chemistry and physics applications from a Monte Carlo perspective},
    journal = {The Journal of Chemical Physics},
    volume = {160},
    number = {1},
    pages = {010901},
    year = {2024},
    month = {01},
    issn = {0021-9606},
    doi = {10.1063/5.0173591},
    url = {https://doi.org/10.1063/5.0173591},
    eprint = {https://pubs.aip.org/aip/jcp/article-pdf/doi/10.1063/5.0173591/19980844/010901_1_5.0173591.pdf},
}

@article{McArdle2020,
  title = {Quantum computational chemistry},
  author = {McArdle, Sam and Endo, Suguru and Aspuru-Guzik, Al\'an and Benjamin, Simon C. and Yuan, Xiao},
  journal = {Rev. Mod. Phys.},
  volume = {92},
  issue = {1},
  pages = {015003},
  numpages = {51},
  year = {2020},
  month = {Mar},
  publisher = {American Physical Society},
  doi = {10.1103/RevModPhys.92.015003},
  url = {https://link.aps.org/doi/10.1103/RevModPhys.92.015003}
}

@article{Motta_2024,
doi = {10.1088/2516-1075/ad3592},
url = {https://doi.org/10.1088/2516-1075/ad3592},
year = {2024},
month = {mar},
publisher = {IOP Publishing},
volume = {6},
number = {1},
pages = {013001},
author = {Motta, Mario and Kirby, William and Liepuoniute, Ieva and Sung, Kevin J and Cohn, Jeffrey and Mezzacapo, Antonio and Klymko, Katherine and Nguyen, Nam and Yoshioka, Nobuyuki and Rice, Julia E},
title = {Subspace methods for electronic structure simulations on quantum computers},
journal = {Electronic Structure}
}

@article{Barthe2025,
  title = {Gate-Based Quantum Simulation of Gaussian Bosonic Circuits on Exponentially Many Modes},
  author = {Barthe, Alice and Cerezo, M. and Sornborger, Andrew T. and Larocca, Mart\'{\i}n and Garc\'{\i}a-Mart\'{\i}n, Diego},
  journal = {Phys. Rev. Lett.},
  volume = {134},
  issue = {7},
  pages = {070604},
  numpages = {6},
  year = {2025},
  month = {Feb},
  publisher = {American Physical Society},
  doi = {10.1103/PhysRevLett.134.070604},
  url = {https://link.aps.org/doi/10.1103/PhysRevLett.134.070604}
}

@misc{D-Wave_Whitepaper,
    title = {Performance gains in the D-Wave Advantage2 system at the
        4,400-qubit scale},
    year = {2025},
    note = {Whitepaper, available at \url{https://www.dwavequantum.com/media/wakjcpsf/adv2_4400q_whitepaper-1.pdf}}
}

@misc{Schuhmacher2025,
      title={Observation of hadron scattering in a lattice gauge theory on a quantum computer}, 
      author={Julian Schuhmacher and Guo-Xian Su and Jesse J. Osborne and Anthony Gandon and Jad C. Halimeh and Ivano Tavernelli},
      year={2025},
      eprint={2505.20387},
      archivePrefix={arXiv},
      primaryClass={quant-ph},
      url={https://arxiv.org/abs/2505.20387}, 
}

@article{Herman2023,
    author = {Herman, Dylan and Googin, Cody andLiu, Xiaoyuan and Sun, Yue and Galda, Alexey and Safro, Ilya and Pistoia, Marco and Alexeev, Yuri},
    title = {Quantum computing for finance},
    journal = {Nature reviews physics},
    year = {2023},
    volume = {5},
    pages = {450-465}
}

@misc{Ciceri2025,
      title={Enhanced fill probability estimates in institutional algorithmic bond trading using statistical learning algorithms with quantum computers}, 
      author={Axel Ciceri and Austin Cottrell and Joshua Freeland and Daniel Fry and Hirotoshi Hirai and Philip Intallura and Hwajung Kang and Chee-Kong Lee and Abhijit Mitra and Kentaro Ohno and Das Pemmaraju and Manuel Proissl and Brian Quanz and Del Rajan and Noriaki Shimada and Kavitha Yograj},
      year={2025},
      eprint={2509.17715},
      archivePrefix={arXiv},
      primaryClass={quant-ph},
      url={https://arxiv.org/abs/2509.17715}, 
}

@article{Shu2024,
    author = {Shu, Guoqiang and Shan, Zheng and Xu, Jinchen and Zhao, Jie and Wang, Shuya},
    title = {A general quantum algorithm for numerical integration},
    journal = {Scientific Reports},
    year = {2024},
    volume = {14}
}

@article{Bochkarev2026,
title = {Quantum computing for discrete optimization: A highlight of three technologies},
journal = {European Journal of Operational Research},
volume = {329},
number = {3},
pages = {747-766},
year = {2026},
issn = {0377-2217},
doi = {https://doi.org/10.1016/j.ejor.2025.07.063},
url = {https://www.sciencedirect.com/science/article/pii/S0377221725005880},
author = {Alexey Bochkarev and Raoul Heese and Sven Jäger and Philine Schiewe and Anita Schöbel}
}

@article{Fellous-Asiani2025,
    author = {Fellous-Asiani, Marco and Naseri, Moein and Datta, Chandan and Streltsov, Alexander and Oszmaniec, Michal},
    title = {Scalable noisy quantum circuits for biased-noise qubits},
    journal = {npj Quantum Information},
    volume = {11},
    issue = {1},
    year = {2025}
}

@article{Yan2025,
    author = {Yan, Yuxuan and Du, Zhenyu and Chen, Junjie and Ma, Xiongfend},
    title = {Limitations of noisy quantum devices in computing and entangling power},
    journal = {npj Quantum Informaiton},
    volume = {11},
    issue = {1},
    year = {2025}
}

@article{Quek2024,
    author = {Quek, Yihui and Stilck Fran\c{c}a, Daniel and Khatri, Sumeet and Meyer, Johannes Jakob and Eisert, Jens},
    title = {Exponentially tighter bounds on limitations of quantum error mitigation},
    journal = {Nature Physics},
    volume = {20},
    issue = {10},
    year = {2024}
}

@article{Roffe2019,
author = {Joschka Roffe},
title = {Quantum error correction: an introductory guide},
journal = {Contemporary Physics},
volume = {60},
number = {3},
pages = {226--245},
year = {2019},
publisher = {Taylor \& Francis},
doi = {10.1080/00107514.2019.1667078},
URL = { 
        https://doi.org/10.1080/00107514.2019.1667078     
},
eprint = {    
        https://doi.org/10.1080/00107514.2019.1667078     
}
}

@inproceedings{Hamilton2020,
title = "Scalable quantum processor noise characterization",
author = "Hamilton, \{Kathleen E.\} and Tyler Kharazi and Titus Morris and McCaskey, \{Alexander J.\} and Bennink, \{Ryan S.\} and Pooser, \{Raphael C.\}",
note = "Publisher Copyright: {\textcopyright} 2020 IEEE.; 2020 IEEE International Conference on Quantum Computing and Engineering, QCE 2020 ; Conference date: 12-10-2020 Through 16-10-2020",
year = "2020",
month = oct,
doi = "10.1109/QCE49297.2020.00060",
language = "English",
series = "Proceedings - IEEE International Conference on Quantum Computing and Engineering, QCE 2020",
publisher = "Institute of Electrical and Electronics Engineers Inc.",
pages = "430--440",
editor = "Muller, \{Hausi A.\} and Greg Byrd and Candace Culhane and Erik DeBenedictis and Travis Humble",
booktitle = "Proceedings - IEEE International Conference on Quantum Computing and Engineering, QCE 2020",

}

@misc{Döbler2024,
      title={Scalable General Error Mitigation for Quantum Circuits}, 
      author={Philip Döbler and Jannik Pflieger and Fengping Jin and Hans De Raedt and Kristel Michielsen and Thomas Lippert and Manpreet Singh Jattana},
      year={2024},
      eprint={2411.07916},
      archivePrefix={arXiv},
      primaryClass={quant-ph},
      url={https://arxiv.org/abs/2411.07916}, 
}

@book{NielsenChuang2010,
    author = {Michael Nielsen and Isaac Chuang},
    title = {Quantum Computation and Quantum Information},
    publisher = {Cambridge University Press},
    year = {2010}
}

@article{Farenick2011,
    author = {Farenick, Douglas and Plosker, Sarah and Smith, Jerrod},
    title = {Classical and nonclassical randomness in quantum measurements},
    journal = {Journal of Mathematical Physics},
    volume = {52},
    number = {12},
    pages = {122204},
    year = {2011},
    month = {12},
    issn = {0022-2488},
    doi = {10.1063/1.3668081},
    url = {https://doi.org/10.1063/1.3668081},
    eprint = {https://pubs.aip.org/aip/jmp/article-pdf/doi/10.1063/1.3668081/15777965/122204_1_online.pdf},
}

@misc{Cedeñopérez2025,
      title={A General Theory of Operator-Valued Measures}, 
      author={Luis A. Cede\~{}o-P\'{e}rez and Hernando Quevedo},
      year={2025},
      eprint={2410.19306},
      archivePrefix={arXiv},
      primaryClass={math.FA},
      url={https://arxiv.org/abs/2410.19306}, 
}

@misc{Jorgensen2025B,
      title={Hilbert space-valued Gaussian processes, and quantum states}, 
      author={Palle E. T. Jorgensen and James Tian},
      year={2025},
      eprint={2405.02796},
      archivePrefix={arXiv},
      primaryClass={quant-ph},
      url={https://arxiv.org/abs/2405.02796}, 
}

@article{Jorgensen2025,
title = {Operator-valued kernels, machine learning, and dynamical systems},
journal = {Physica D: Nonlinear Phenomena},
volume = {476},
pages = {134657},
year = {2025},
issn = {0167-2789},
doi = {https://doi.org/10.1016/j.physd.2025.134657},
url = {https://www.sciencedirect.com/science/article/pii/S0167278925001368},
author = {Palle E.T. Jorgensen and James Tian}
}

@article{Geller2013,
  title = {Efficient error models for fault-tolerant architectures and the Pauli twirling approximation},
  author = {Geller, Michael R. and Zhou, Zhongyuan},
  journal = {Phys. Rev. A},
  volume = {88},
  issue = {1},
  pages = {012314},
  numpages = {7},
  year = {2013},
  month = {Jul},
  publisher = {American Physical Society},
  doi = {10.1103/PhysRevA.88.012314},
  url = {https://link.aps.org/doi/10.1103/PhysRevA.88.012314}
}

@misc{Linh2025,
      title={Advancing quantum process tomography through universal compilation}, 
      author={Huynh Le Dan Linh and Vu Tuan Hai and Le Bin Ho},
      year={2025},
      eprint={2504.14958},
      archivePrefix={arXiv},
      primaryClass={quant-ph},
      url={https://arxiv.org/abs/2504.14958}, 
}

@article{Flammia2012,
    author = {Flammia, Steven T and Gross, David and Liu, Yi-Kai and Eisert, Jans},
    title = {Quantum tomography via compressed sensing: error bounds, sample complexity, and efficient estimators},
    journal = {New Journal of Physics},
    volume = {14},
    year = {2012}
}

@article{AbuGhanem2025,
    author = {Muhammad, AbuGhanem},
    title = {Full {Q}uantum {P}rocess {T}omography of a {U}niversal {E}ntangling {G}ate on an {IBM}'s {Q}uantum {C}omputer},
    journal = {Arabian Journal for Science and Enginnering},
    volume = {50},
    issue = {23},
    year = {2025}
}

@article{Torlai2023,
    author = {Torlai, Giacomo and Wood, Christoper J. and Acharya, Atithi and Carleo, Giuseppe and Carrasquilla, Juan and Aolita, Leandro},
    title = {Quantum process tomography with unsupervised learning and tensor networks},
    journal = {Nature Communications},
    volume = {14},
    year = {2023}
}

@article{Hosny2023,
title = {Survey on compressed sensing over the past two decades},
journal = {Memories - Materials, Devices, Circuits and Systems},
volume = {4},
pages = {100060},
year = {2023},
issn = {2773-0646},
doi = {https://doi.org/10.1016/j.memori.2023.100060},
url = {https://www.sciencedirect.com/science/article/pii/S2773064623000373},
author = {Sherif Hosny and M. Watheq El-Kharashi and Amr T. Abdel-Hamid}
}

@article{Nagahara2023,
    author = {Nagahara, Masaaki and Yamamoto, Yutaka},
    title = {A survey on compressed sensing approach to systems and control},
    journal = {Mathematics of Control, Signals, and Systems},
    volume = {36},
    year = {2024}
}

@Inbook{Altepeter2004,
author="Altepeter, Joseph B.
and James, Daniel F.V.
and Kwiat, Paul G.",
editor="Paris, Matteo
and {\v{R}}eh{\'a}{\v{c}}ek, Jaroslav",
title="4 Qubit Quantum State Tomography",
bookTitle="Quantum State Estimation",
year="2004",
publisher="Springer Berlin Heidelberg",
address="Berlin, Heidelberg",
pages="113--145",
isbn="978-3-540-44481-7",
doi="10.1007/978-3-540-44481-7_4",
url="https://doi.org/10.1007/978-3-540-44481-7_4"
}

@article{Vogel1989,
  title = {Determination of quasiprobability distributions in terms of probability distributions for the rotated quadrature phase},
  author = {Vogel, K. and Risken, H.},
  journal = {Phys. Rev. A},
  volume = {40},
  issue = {5},
  pages = {2847--2849},
  numpages = {0},
  year = {1989},
  month = {Sep},
  publisher = {American Physical Society},
  doi = {10.1103/PhysRevA.40.2847},
  url = {https://link.aps.org/doi/10.1103/PhysRevA.40.2847}
}

@article{Rehacek2001,
  title = {Iterative algorithm for reconstruction of entangled states},
  author = {\ifmmode \check{R}\else \v{R}\fi{}eh\'a\ifmmode \check{c}\else \v{c}\fi{}ek, J. and Hradil, Z. and Je\ifmmode \check{z}\else \v{z}\fi{}ek, M.},
  journal = {Phys. Rev. A},
  volume = {63},
  issue = {4},
  pages = {040303},
  numpages = {4},
  year = {2001},
  month = {Mar},
  publisher = {American Physical Society},
  doi = {10.1103/PhysRevA.63.040303},
  url = {https://link.aps.org/doi/10.1103/PhysRevA.63.040303}
}

@misc{Lum2022,
      title={Bayesian Quantum State Tomography with Python's PyMC}, 
      author={Daniel J. Lum and Yaakov Weinstein},
      year={2022},
      eprint={2212.10655},
      archivePrefix={arXiv},
      primaryClass={quant-ph},
      url={https://arxiv.org/abs/2212.10655}, 
}

@article{Goncalves2011,
  title={Local solutions of maximum likelihood estimation in quantum state tomography},
  author={Douglas Soares Gonçalves and M{\'a}rcia A. Gomes-Ruggiero and Carlile Lavor and Osvaldo Jim{\'e}nez Farias and Paulo H. Souto Ribeiro},
  journal={Quantum Inf. Comput.},
  year={2011},
  volume={12},
  pages={775-790},
  url={https://api.semanticscholar.org/CorpusID:6550190}
}

@article{Innan2024,
    author = {Innan, Nouhaila and Siddiqui, Owais Ishtiaq and Arora, Shivang and Ghogh, Tamojit and Ko\c{c}ak, Yasemin Poyraz and Paragas, Dominic and Galib, Abdullah Al Omar and Khan, Muhammad Al-Zafar and Bennai, Mohamed},
    title = {Quantum state tomography using quantum machine learning},
    journal = {Quantum Machine Intelligence},
    volume = 6,
    year = {2024}
}

@article{Wu2025,
    author = {Wu, Hsun-Chung and Hsieh, Hsien-Yi and Xu, Zhi-Kai and Chen, Hua Li and Shi, Zi-Hao and Wang, Po-Han and Yang, Popo and Steuernagel, Ole and Suen, Te-Hwei and Wu, Chien-Ming and Lee, Ray-Kuang},
    title = {Machine learning enhanced quantum state tomography on a field-programmable gate array},
    journal = {APL Quantum},
    volume = {2},
    number = {2},
    pages = {026117},
    year = {2025},
    month = {05},
    issn = {2835-0103},
    doi = {10.1063/5.0262942},
    url = {https://doi.org/10.1063/5.0262942},
    eprint = {https://pubs.aip.org/aip/apq/article-pdf/doi/10.1063/5.0262942/20518411/026117_1_5.0262942.pdf},
}

@article{Leung2002,
  title={Choi’s proof as a recipe for quantum process tomography},
  author={Debbie W. Leung},
  journal={Journal of Mathematical Physics},
  year={2002},
  volume={44},
  pages={528-533},
  url={https://api.semanticscholar.org/CorpusID:14895994}
}

@misc{Greenbaum2015,
      title={Introduction to Quantum Gate Set Tomography}, 
      author={Daniel Greenbaum},
      year={2015},
      eprint={1509.02921},
      archivePrefix={arXiv},
      primaryClass={quant-ph},
      url={https://arxiv.org/abs/1509.02921}, 
}

@article{Nielsen2021,
  doi = {10.22331/q-2021-10-05-557},
  url = {https://doi.org/10.22331/q-2021-10-05-557},
  title = {Gate {S}et {T}omography},
  author = {Nielsen, Erik and Gamble, John King and Rudinger, Kenneth and Scholten, Travis and Young, Kevin and Blume-Kohout, Robin},
  journal = {{Quantum}},
  issn = {2521-327X},
  publisher = {{Verein zur F{\"{o}}rderung des Open Access Publizierens in den Quantenwissenschaften}},
  volume = {5},
  pages = {557},
  month = oct,
  year = {2021}
}

@article{Cattaneo2023,
  title = {Self-consistent quantum measurement tomography based on semidefinite programming},
  author = {Cattaneo, Marco and Rossi, Matteo A. C. and Korhonen, Keijo and Borrelli, Elsi-Mari and Garc\'{\i}a-P\'erez, Guillermo and Zimbor\'as, Zolt\'an and Cavalcanti, Daniel},
  journal = {Phys. Rev. Res.},
  volume = {5},
  issue = {3},
  pages = {033154},
  numpages = {14},
  year = {2023},
  month = {Sep},
  publisher = {American Physical Society},
  doi = {10.1103/PhysRevResearch.5.033154},
  url = {https://link.aps.org/doi/10.1103/PhysRevResearch.5.033154}
}

@article{Kuegler2012,
author = {Kuegler, Philipp},
title = {A Sparse Update Method for Solving Underdetermined Systems of Nonlinear Equations Applied to the Manipulation of Biological Signaling Pathways},
journal = {SIAM Journal on Applied Mathematics},
volume = {72},
number = {4},
pages = {982-1001},
year = {2012},
doi = {10.1137/110834780},

URL = { 
    
        https://doi.org/10.1137/110834780
    
    

},
eprint = { 
    
        https://doi.org/10.1137/110834780
    
    

}
}

@article{Cline1976,
author = {Cline, R. E. and Plemmons, R. J.},
title = {\$l\_2 \$-Solutions to Underdetermined Linear Systems},
journal = {SIAM Review},
volume = {18},
number = {1},
pages = {92-106},
year = {1976},
doi = {10.1137/1018004},

URL = { 
    
        https://doi.org/10.1137/1018004
    
    

},
eprint = { 
    
        https://doi.org/10.1137/1018004
    
    

}
}

@INPROCEEDINGS{Donoho2006,
  author={Donoho, David and Kakavand, Hossein and Mammen, James},
  booktitle={2006 IEEE International Symposium on Information Theory}, 
  title={The Simplest Solution to an Underdetermined System of Linear Equations}, 
  year={2006},
  volume={},
  number={},
  pages={1924-1928},
  doi={10.1109/ISIT.2006.261816}}

@article{Lines1984,
author = {LINES, L.R. and TREITEL, S.},
title = {A REVIEW OF LEAST-SQUARES INVERSION AND ITS APPLICATION TO GEOPHYSICAL PROBLEMS},
journal = {Geophysical Prospecting},
volume = {32},
number = {2},
pages = {159-186},
doi = {https://doi.org/10.1111/j.1365-2478.1984.tb00726.x},
url = {https://onlinelibrary.wiley.com/doi/abs/10.1111/j.1365-2478.1984.tb00726.x},
eprint = {https://onlinelibrary.wiley.com/doi/pdf/10.1111/j.1365-2478.1984.tb00726.x},
year = {1984}
}

@article{Opfell1958,
    author = {Opfell, J.B. and Sage, B.H.},
    title = {Applications of Least Squares Methods},
    journal = {Industrial Engineering Chemistry},
    volume = {50},
    issue = {5},
    year = {1958}
}

@article{Galindo-Prieto2026,
title = {A review of the applications and challenges of applying partial least squares (PLS) to exposomics research},
journal = {Analytica Chimica Acta},
volume = {1383},
pages = {344826},
year = {2026},
issn = {0003-2670},
doi = {https://doi.org/10.1016/j.aca.2025.344826},
url = {https://www.sciencedirect.com/science/article/pii/S0003267025012206},
author = {Beatriz Galindo-Prieto and Ian S. Mudway}
}

@article{Miller2025,
author = {Steven J. Miller},
title = {The Method of Least Squares: The Theory and Applications of Linear Regression, from the Orbits of Planets to Regionalizing School Districts},
journal = {Scatterplot},
volume = {2},
number = {1},
pages = {2470533},
year = {2025},
publisher = {Taylor \& Francis},
doi = {10.1080/29932955.2025.2470533},


URL = { 
    
        https://doi.org/10.1080/29932955.2025.2470533
    
    

},
eprint = { 
    
        https://doi.org/10.1080/29932955.2025.2470533
    
    

}

}

@article{Hyun2021,
title = {Deep learning-based solvability of underdetermined inverse problems in medical imaging},
journal = {Medical Image Analysis},
volume = {69},
pages = {101967},
year = {2021},
issn = {1361-8415},
doi = {https://doi.org/10.1016/j.media.2021.101967},
url = {https://www.sciencedirect.com/science/article/pii/S136184152100013X},
author = {Chang Min Hyun and Seong Hyeon Baek and Mingyu Lee and Sung Min Lee and Jin Keun Seo}
}

@misc{Bertalan2025,
      title={Data-Driven, ML-assisted Approaches to Problem Well-Posedness}, 
      author={Tom Bertalan and George A. Kevrekidis and Eleni D Koronaki and Siddhartha Mishra and Elizaveta Rebrova and Yannis G. Kevrekidis},
      year={2025},
      eprint={2503.19255},
      archivePrefix={arXiv},
      primaryClass={cs.LG},
      url={https://arxiv.org/abs/2503.19255}, 
}

@misc{xu2025efficientmeasurementerrormitigation,
      title={Efficient Measurement Error Mitigation with Subsystem-Balanced Pauli Twirling}, 
      author={Xiao-Yue Xu and Chen Ding and Wan-Su Bao},
      year={2025},
      eprint={2509.17298},
      archivePrefix={arXiv},
      primaryClass={quant-ph},
      url={https://arxiv.org/abs/2509.17298}, 
}

\end{document}